\documentstyle[preprint,revtex]{aps}
\begin{document}
\draft
\begin{title}
Pion interaction with the trinucleon up to the eta production threshold
\end{title}
\author{S. S. Kamalov\cite{Sabit} and L. Tiator}
\begin{instit}
Institut f\"ur Kernphysik, Universit\"at Mainz, 6500 Mainz, Germany
\end{instit}
\author{C. Bennhold}
\begin{instit}
TRIUMF, 4004 Wesbrook Mall, Vancouver, B.C. V6T 2A3, Canada
\end{instit}
\begin{abstract}
Pion elastic, charge exchange scattering and induced eta production on the
trinucleon systems are investigated in a coupled-channels approach in momentum
space with Fadeev wave functions. The channel $\pi N \rightarrow \eta N$
is included using an isobar model with S-, P-, and D-wave resonances.
While the coherent reactions like $^3$He($\pi,\pi)^3$He can be reasonably
well reproduced up to $T_{\pi}$=500 MeV, large discrepancies appear for the
incoherent processes, $^3$He($\pi^-,\pi^0)^3$H and $^3$He($\pi^-,\eta)^3$H at
backward angles and energies above $\Delta$-resonance.
In the forward direction the $(\pi,\eta)$ calculations underestimate the
experimental measurements very close to threshold but agreement with the data
improves with increasing pion energy.
Predictions are made for the asymmetries of the various
reactions on polarized $^3$He.
\end{abstract}


\newpage

\section{INTRODUCTION}

For many years there has been a great interest in understanding the
pion-nuclear interaction at low and intermediate energies. A recent
review of achievements in this field can be found in Ref.\cite{1}.
Originally, the phenomenological Kisslinger potential in coordinate space
was derived for spin and isospin  zero nuclei only as a first step.
 After that  numerous
modifications and improvements were suggested to correct the optical potential:
real pion absorption, Fermi motion, Lorentz-Lorentz effects, Pauli blocking
effects etc.

On the other hand, a microscopic description of the pion-nuclear interaction
in momentum space was developed based on the KMT\cite{2} or Watson\cite{3}
multiple scattering theories. In the framework of such an approach the
nonlocalities of the pion-nuclear interaction, off-shell extrapolations of
pion-nucleon scattering amplitudes and exact treatment of Fermi motion have
been taken into account\cite{4,5}. In a further step the phenomenological
$\rho^2$ term has been added\cite{6,7,8} which is responsible for real pion
absorption and second order effects.
The momentum space formalism was not only successful in the description of the
pion-nuclear interaction in the
$\Delta$ resonance region but also at low energies
for a large set of nuclei with A=4--40. In Refs.\cite{9,10} this method
has been extended successfully to the description of pionic atoms
as well.

  At the present time studies of the pion interaction with the trinucleon
system have become very attractive. First, this is due to the development of
new experimental techniques at TRIUMF, PSI and LAMPF which allow measurements
of polarization observables. Secondly, one already has a large set of
previous measurements for the differential cross section. On the theoretical
side the nuclear structure of the trinucleon system is well known . All these
circumstances create ideal conditions to study in detail the pion interaction
mechanisms with very light nuclei.

Previous theoretical investigations of the pion interaction with
the trinucleon system were mostly based on the multiple scattering theory and,
according to a recent analysis by Gibbs and Gibson\cite{11}, the KMT version
of this approach is preferable. In this framework Landau\cite{4} and
Mach\cite{6} studied the sensitivity of pion elastic scattering and single
charge exchange (SCE)on $^3$He and $^3$H to the details of the nuclear wave
function. While Mach used semiphenomenological wave functions, Landau
extracted the four nuclear form factors required for the optical potential
from the electromagnetic form factors of $^3$He and $^3$H.
Clearly these two approaches differ since the parametrized form factors
contain meson exchange current (MEC) contributions. Nevertheless, both
achieve a good description of experimental data up to pion kinetic
energy of $T_{\pi}$=200 MeV.

At higher energies large discrepancies appear between the measurements
and theoretical calculations of elastic scattering and SCE at backward angles.
For example, in the SCE reaction at $T_{\pi}$=300 MeV and at pion angles
$\Theta>90$$^{\circ}$ basically all calculations yield differential cross
sections which fall two orders of magnitude below the data. Only
calculations in the framework of the Glauber approach\cite{12,13} reached
better agreement in this region. However, we concur with Ref.\cite{14})
that in the large-angle region Glauber theory cannot be reliable.

The solution to this problem could involve corrections
to the impulse and coherent approximation in form of second order effects.
Several attempts to incorporate such corrections for the pion interaction
with very light nuclei are discussed in Refs.\cite{14,15,16,17}. A full
analysis of second order effects including spin and isospin degrees of
freedom has been  performed by Wakamatsu\cite{14}. According to his work
these contributions are larger in pion SCE than in elastic scattering
and they can enhance the differential cross section at backward angles
by a factor of two. This, however, is not sufficient to explain the data.

Another approach was developed by Gibbs and collaborators\cite{11,18}.
Here the $s$- and $p$-wave parts  of the first-order potential were
calculated using the frozen nucleon approximation, then the obtained
potential was averaged over realistic nuclear configurations. In such an
approach part of the second-order effects are automatically included via
the $NN$ correlations in the nuclear wave function. Refs.\cite{11,18} yield
good agreement for pion energies $T_{\pi}$=24--180 MeV, however, applying this
approach to higher energies where higher pion-nucleon partial waves become
important encounters computational difficulties.

The aim of the present work is a systematic investigation of the pion
interaction  with the trinucleon systems ( $^3$He and $^3$H) in the region of
pion kinetic energies of $T_{\pi}$=100--600 MeV, including spin and isospin
degrees of freedom on the microscopic level and
using only elementary amplitudes extracted from $\pi N$ scattering data
and realistic three body wave functions obtained as a solution of the Fadeev
equations. Due to the last circumstance we are in a very good position to fix
the nuclear structure input in order to shed more light on the reaction
mechanism. In the framework of such an approach in our previous studies
of pion scattering on unpolarized and polarized
$^3$He-targets\cite{19,20} we have reached a good description of older and
recent measurements at low pion energies. This has motivated us to extend our
momentum space approach for  pion energies up to the eta production region.
This is the kinematical region $T_{\pi}< 600$ MeV
and momentum transfer $Q < 6 fm^{-1}$, where Fadeev wave functions give an
excellent description of the electromagnetic form factors in the  A=3 system.
This gives us confidence to study the reaction mechanism. In this kinematical
region we will check our model by comparison with all available experimental
data and present predictions for forthcoming experiments at LAMPF, PSI and
TRIUMF.

The main aspects of our formalism based on the KMT multiple scattering
approach and coupled-channels method are given in section 2.
In section 3 we describe the nuclear models: 1) a simple S-shell
model which is very useful for a qualitative (and in some kinematical regions
even quantitative) description of the main features of the pion interaction
with the trinucleon\cite{20}; 2) a phenomenological model employing
nuclear form factors extracted from electron scattering data which was often
used in previous calculations; 3) a three-body model with realistic wave
functions obtained from solutions of the Fadeev equations\cite{21}.
The aim of this comparison  with different nuclear models
is to estimate the upper limit of uncertainty which was present in earlier
investigations of pion interaction with $^3$He.

In section 4, our results for elastic scattering and single charge
exchange on unpolarized and polarized targets are discussed. Furthermore,
we briefly consider the "superratio" which has been used\cite{11} to find
charge symmetry breaking in strong interactions.

We will complete our analysis with the investigation of eta production by
pions.
This is a relatively new field of theoretical\cite{22,23,24} and experimental
\cite{25} study in pionic physics which promises to be very useful
to obtain a global picture  of the meson-nuclear interaction. On the other
hand, these studies could give very important information on the excitation of
nucleon resonances in nuclei (like the $S_{11}$(1535) and the $D_{13}$(1525))
that could be applied to other nuclear reactions with eta mesons (i.e.,
eta production in heavy-ion collisions and eta photoproduction )

Our results and conclusions are summarized in section 5 and we will give
the main expressions for the pion-nucleon and eta-nucleon interactions
in the Appendix.

\section{FORMALISM}

\subsection{General expressions}

The general formalism for the description of the pion-nuclear scattering
is based on a momentum space coupled-channels method which was developed
in detail in Ref.\cite{26}. Here we merely summarize its main features.

In the framework of the KMT version of multiple scattering theory, the
scattering amplitude is constructed by solving the Lippmann-Schvinger integral
equation with relativistic kinematics:
\begin{equation}
{\rm F}_{\pi' \pi}(\vec{q}\,',\vec{q}) ={\rm V}_{\pi' \pi}
(\vec{q}\,',\vec{q})-\frac{a}{(2\pi)^2} \sum_{\pi''} \int \frac{d\vec{q}\,''}
{{\cal{M}}(q'')}\frac{{\rm V}_{\pi'\pi''}(\vec{q}\,',\vec{q}\,'')
{\rm F}_{\pi''\pi}(\vec{q}\,'',\vec{q})}{E(q) - E(q'') + i\epsilon} \, \, ,
\end{equation}
where $\vec{q}$ is the pion momentum, and
 $E(q)=E_{\pi}(q)+E_{A}(q)$ is the total
pion-nuclear energy. The pion-nuclear reduced mass is given by
${\cal M}(q)=E_{\pi}(q)E_{A}(q)/E(q)$ and the coefficient $a=(A-1)/A$ is
important to avoid double counting.

The momentum space potential of pion-nuclear interaction
\begin{equation}
{\rm V}_{\pi' \pi}(\vec{q}\,',\vec{q})=
{\rm V}_{{\rm {Coul.}}}(\vec{q}\,'-\vec{q};R)+{\rm V}_{\pi' \pi}^{(1)}
(\vec{q}\,',\vec{q})
\end{equation}
contains the Coulomb potential in the momentum space, cut at point R,
and a potential of strong pion-nuclear interaction ${\rm V}_{\pi' \pi}^{(1)}$
 which is related to the free $\pi N$ scattering t-matrix:
\begin{equation}
{\rm V}_{\pi' \pi}^{(1)}(\vec{q}\,',\vec{q})=
-\frac{\sqrt{{\cal M}(q'){\cal M}(q)}}{2\pi}<\pi' (\vec{q}\,'),f \mid
\sum^A_{j=1} \hat{\rm t}_{\pi N}(j)\mid i,\pi (\vec{q})>\,,
\end{equation}
where $\mid i>$ and $\mid f>$ denote the nuclear initial and final
states, respectively, and  $j$ refers to the individual target nucleons.

The pion-nucleon scattering t-matrix is defined in the following way
\begin{equation}
\hat{\rm t}_{\pi N}=-\frac{2\pi}{\sqrt{\mu (q',p')\mu (q,p)}}
\left[ {A_{00}+A_{01}\vec{t}\cdot\vec{\tau}+i\vec{\sigma}\cdot
[\hat{\vec{q}}_{i}\times\hat{\vec{q}}_{f}]
(A_{10}+A_{11}\vec{t}\cdot\vec{\tau})}\right]\,,
\end{equation}
where $\mu (q,p)=E_{\pi}(q)E_{N}(p)/\omega$ is the pion-nucleon reduced mass,
$\vec{p}$ and $\vec{p}\,'=\vec{p}+\vec{q}-\vec{q}\,'$ are the nucleon momenta
in the initial and final states, respectively, and $\hat{\vec{q}}_{i}$ and
$\hat{\vec{q}}_{f}$ are the unit vectors for the initial and final pion
momentum in the $\pi N$ c.m. system.  The vectors
$\vec{\sigma}$, $\vec{\tau}$ and
$\vec{t}$ are the usual spin and isospin operators of target nucleon and pion,
respectively.

The scalar functions $A_{ST}(\omega,\cos{\Theta^*})\,(S=0,1$ and $T=0,1$),
which depend on the total pion-nucleon energy $\omega$ and the pion angle
$\Theta^*$ in the $\pi N$ c.m. system, are the usual combinations of partial
$\pi N$ scattering amplitudes $f_{l_\pi}^{(\pm)}(\omega)$ and Legendre
polynomials $P_{l_\pi}(\cos{\Theta^*})$  where $l_\pi$ is the pion-nucleon
angular momentum and $(\pm)$ corresponds to the total spin $j_\pi=l_\pi \pm
\frac{1}{2}$ of the $\pi N$ system. These amplitudes are constructed from
the pion-nucleon phase-shift parametrization of ref.\cite{27} at
$m_{\pi}+M_{N}<\omega<1.3$ GeV and the CERNTH-parametrization\cite{28} at
$1.3 {\rm GeV}<\omega<2.2$ GeV.

The discussion of all problems connected with the partial wave analysis of
different groups is beyond this paper. A review of this subject can be found
in the last compilation of\cite{29}.
Note only that in the kinematical region considered here ($\omega<1.6$ GeV),
the difference between the results of partial wave analysis of different
groups is small (much smaller than the accuracy of the nuclear data).
For example, a comparison of CERNTH and VPI\cite{30} analyses show no
difference in the P33 wave up to $w=1.6$ GeV. Also $P11$ and $D13$ phase
shifts are very similar in this region. Only for $S11$ the difference
between results of these two groups becomes sizable at $w>1.5$ GeV\cite{22},
i.e. in the region were the eta channel is open. Note that recently
a new $\pi N$ scattering partial analysis was published by the VPI group
\cite{31} which  gives a new determination of $S11$ resonance parameters
consistent with results of other groups.

The off-shell extrapolation of $\pi N$ parcial amplitudes was obtained using
the separable form
\begin{equation}
f_{l_\pi}^{(\pm)}({\rm {off-shell}})=f_{l_\pi}^{(\pm)}(\omega)
\frac{v_{l_\pi}^{(\pm)}(q'') v_{l_\pi}^{(\pm)}(q')}{[v_{l_\pi}^{(\pm)}(q)]^2}
\end{equation}
with $\pi N$ form factors
\begin{equation}
v_{l_\pi}^{(\pm)}(q) = \frac{q^{l_\pi}}{[1+(r_{0}q)^2]^2}\,.
\end{equation}
Here we employ the value $r_0 = 0.47 fm$ consistent with the analysis
of the separable $\pi N$ potential in Ref.\cite{32}.

To connect the total pion-nucleon energy $\omega$, at which these partial
amplitudes are  calculated, with the total pion-nuclear energy $E$ we will
use the results of a relativistic generalization for the three-body model
\cite{6,26}:
\begin{eqnarray}
\omega & = & E+m_\pi+M_N- \left\{ (m_\pi+M_N)^2+
\left[ \frac{A-1}{2A}(\vec{q}\,'+\vec{q})^2 \right]^2 \right\}^{\frac{1}{2}}-
\nonumber \\ & &
-\left\{(A-1)^2 M_N^2+\left[\frac{A-1}{2A}(\vec{q}\,'+\vec{q})^2 \right]^2
\right\}^{\frac{1}{2}}\,.
\end{eqnarray}
Such an approach allowed us to improve the impulse approximation in accordance
with the results of Ref.\cite{33}.

\subsection{Partial wave decomposition and the amplitudes
 ${\cal F}$ and ${\cal G}$ }

In our numerical applications we express the scattering amplitude in terms
of partial amplitudes using  the representations of total isospin $I$
and projection $\xi$
\begin{equation}
{\rm F}_{\pi' \pi}(\vec{q}\,',\vec{q})=\sum_{I,\xi}(2I+1)
\left( {1\;\;\,\frac{1}{2}\;\;\;\;I}
\atop{\pi\;\;\nu_i\;\;{\rm -}\xi} \right)
\left( {1\;\;\,\frac{1}{2}\;\;\;\;\,I}
\atop{\pi'\;\;\nu_f\;\;{\rm -}\xi} \right)
{\rm F}_{I}(\vec{q}\,',\vec{q})
\end{equation}
and of total angular momentum $j$ with projection $m$
\begin{equation}
{\rm F}_{I}(\vec{q}\,',\vec{q})\!=\!4\pi\sum (2j+1)
Y_{M'_{\pi}}^{L_{\pi}}(\hat{\vec{q'}}) F^{j,I}_{L_{\pi}}
(q',q)Y_{M_{\pi}}^{*L_{\pi}}(\hat{\vec{q}})\!
\left({L_{\pi}\;\;\frac{1}{2}\;\;\;\;\;j}
\atop{M'_{\pi}\,M_f\,{\rm -}m}\right)\!
\left({L_{\pi}\;\;\frac{1}{2}\;\;\;\;j}
\atop{M_{\pi}\,M_i\,{\rm -}m}\right)\,,\;\;\;\;
\end{equation}
where $\left({.\,.\,.}\atop{.\,.\,.}\right)$ stand for the $3j$ symbol,
$\nu_i$, $\nu_f$ are the isospin projectors of the initial and final nucleus,
$Y_{M_{\pi}}^{L_{\pi}}(\hat{\vec{q}})$ are spherical harmonics for the
pion waves and $L_{\pi}$ is the pion angular momentum. Note that in the case
of isoelastic pion scattering on the A=3 system the pion
orbital angular momentum in the initial and final states are equal due to
parity conservation. For the potential ${\rm V}_{\pi' \pi}$ we perform
an expansion identical to eqs.(8,9).

Substituting the above expansions of ${\rm V}_{\pi' \pi}$ and
${\rm F}_{\pi' \pi}$ into eq.(1) we obtain the following integral equation for
the partial wave amplitudes
\begin{equation}
F^{j,I}_{L_{\pi}}(q',q)=V^{j,I}_{L_{\pi}}(q',q)-\frac{a}{\pi} \int
\frac{q''^2}{{\cal{M}}(q'')}\frac{V^{j,I}_{L_{\pi}}(q',q'')
F^{j,I}_{L_{\pi}}(q'',q)}{E(q) - E(q'') + i\epsilon} dq''\,.
\end{equation}
This equation is solved using the matrix inversion method. To obtain
the final values for the partial amplitudes we also take into account
the Coulomb interaction applying the matching procedure developed
by Vincent and Phatak\cite{34}.

Now let us consider some isospin symmetry properties that follow from eq.(8)
when the Coulomb force is turned off. Overall, this is a good
approximation except for the small angle region. In this isospin symmetric
case for the A=3 system we have the following relations:\\
for the $\pi^{\pm}$ elastic scattering amplitudes
\begin{mathletters}
\begin{equation}
<\pi^{+},\,^3{\rm He} \mid {\rm F} \mid\,^3{\rm He},\,\pi^{+}> =
<\pi^{-},\,^3{\rm H} \mid{\rm F} \mid\,^3{\rm H},\,\pi^{-}> =
{\rm F}_{\frac{3}{2}}
\end{equation}
\begin{equation}
<\pi^{-},\,^3{\rm He} \mid {\rm F} \mid\,^3{\rm He},\,\pi^{-}> =
<\pi^{+},\,^3{\rm H} \mid {\rm F} \mid\,^3{\rm H},\,\pi^{+}> =
\frac{1}{3}({\rm F}_{\frac{3}{2}}+2{\rm F}_{\frac{1}{2}})
\end{equation}
and for the single charge exchange reaction
\begin{equation}
<\pi^{0},\,^3{\rm H} \mid {\rm F} \mid\,^3{\rm He},\,\pi^{-}> =
<\pi^{0},\,^3{\rm He} \mid{\rm F} \mid\,^3{\rm H},\,\pi^{+}> =
\frac{\sqrt{2}}{3}({\rm F}_{\frac{3}{2}}-{\rm F}_{\frac{1}{2}})\,.
\end{equation}
\end{mathletters}

{}From relations (11) it follows that the pion interaction with the A=3 systems
is described by the independent amplitudes $F_{\frac{1}{2}}$ and
$F_{\frac{3}{2}}$ which correspond to scattering in the channels with
total pion-nuclear isospins $I=\frac{1}{2}$ and $\frac{3}{2}$, respectively.
In general, these amplitudes are complex.  Moreover, the inclusion of spin
degrees of freedom leads to their additional decomposition into non spin-flip
and spin-flip parts. Therefore, studying only differential cross sections for
elastic scattering and the SCE reactions is not enough to obtain full
information about these amplitudes.

Using the spin structure of the free $\pi N$ scattering amplitude (see eq.(4))
the isotopic ${\rm F}_{I}$ amplitude can be represented as
\begin{equation}
{\rm F}_{I}(q,\cos{\Theta}) = {\cal{F}}_{I}(q,\cos{\Theta}) +
i\,{\cal{G}}_{I}(q,\cos{\Theta})\, \sin{\Theta}\, \vec{\sigma}\cdot
\hat{n}\,.
\end{equation}
Here we assumed (in accordance with the Madison convention) that the
incident pion
momentum $\vec{q}$ is along the positive z-axis and the vector
$\hat{n}=(\vec{q} \times \vec{q}\,')/\mid \vec{q} \times \vec{q}\,'\mid$
is along the positive y-axis in the right-handed coordinate system.

In the plane wave impulse approximation the amplitudes ${\cal{F}}_{I}$ and
${\cal{G}}_{I}$ are directly connected with the non spin-flip
($A_{00},A_{01}$) and spin-flip ($A_{10},A_{11}$) parts of the
elementary amplitude of eq.(4) (see also next section). But taking into
account the contributions of pion rescattering by solving the integral
equation (10) leads to a more complicated situation. In this case both
non spin-flip and spin-flip parts of the elementary amplitude give
contributions to ${\cal{F}}_{I}$ and ${\cal{G}}_{I}$.

Expressions for the total amplitudes ${\cal{F}}$ and ${\cal{G}}$ can be
obtained by summing the partial amplitudes in eqs.(9) and (10).
They are the same as in $\pi N$ scattering, i.e.
\begin{mathletters}
\begin{equation}
{\cal{F}}={\cal F}_{\rm Coul.}+\sum_{L_{\pi}}\exp{(2i\delta_{L_\pi}^C)}
[(L_{\pi}+1)F_{L_{\pi}}^{(+)}(q) + L_{\pi}F_{L_{\pi}}^{(-)}(q)]
P_{L_{\pi}}(\cos{\Theta})
\end{equation}
\begin{equation}
{\cal{G}}=\sum_{L_{\pi}}\exp{(2i\delta_{L_\pi}^C)}[F_{L_{\pi}}^{(+)}(q) -
F_{L_{\pi}}^{(-)}(q)] P'_{L_{\pi}}(\cos{\Theta})\,,
\end{equation}
\end{mathletters}
where $\delta_{L_\pi}^C$ is the point Coulomb phase shift, $P_{L}$ and
$P\,'_{L}$ are the Legendre polynomial and its derivative, respectively, and
$(\pm)$ corresponds to the total spin $j=L_{\pi} \pm \frac{1}{2}$ of the
$\pi A$ system. For the point Coulomb phase shifts $\delta_{L_\pi}^C$ and the
amplitude ${\cal F}_{\rm {Coul}.}$ we used standard expressions given in
Ref.\cite{34}

The differential cross section gives information  on the incoherent sum of
${\cal F}$ and ${\cal G}$ , namely
\begin{equation}
\frac{d\sigma}{d\Omega} = \mid {\cal F} \mid^2 + \mid {\cal G}\mid^{2}
\,\sin^2{\Theta} \, \, .
\end{equation}
On the other hand, polarization observables offer the possibility of learning
more about these two complex angle-dependent functions. For example, the
asymmetry observable depends on their interference
\begin{equation}
 A_y = \frac{2Im({\cal FG}^*)\,\sin{\Theta}}{\mid {\cal F}\mid^2 +
\mid {\cal G} \mid^{2} \sin^2{\Theta}} \, \, .
\end{equation}
The real part could in principle be measured in a double polarization
experiment by detecting the recoil polarization of the final nucleus after
pion scattering from a polarized target.

The full analysis of differential cross sections and
polarization observables for all the reactions listed in eq.(11) can give
us complete information about the spin and isospin parts of the
pion interaction with the A=3 system. Thus we have the opportunity  to test
their symmetry properties and to obtain new information about the nature of
the pion-nuclear interaction.

\section{NUCLEAR MODELS}

\subsection{The three-body model}

Now we turn to the potential ${\rm V}_{\pi' \pi}$ defined in
eqs.(3,4). For the trinucleon system the nuclear wave
functions which enter this expression have to be given by Fadeev
calculations with realistic nucleon-nucleon potentials. In our calculations
we use a wave function which has been obtained in Ref.\cite{21}
with the Reid soft-core potential. This wave function describes both static
and dynamical properties of the A=3 system at momentum transfers
$0<Q<5-6 fm^{-1}$, provided one includes meson exchange
currents in electromagnetic observables.

Using this three-body wave function we can
rewrite the expression (3) for the  potential ${\rm V}_{\pi' \pi}$ in the
following way \cite{35},
\begin{equation}
{\rm V}_{\pi' \pi}(\vec{q}\,',\vec{q}) = -3 \frac{\sqrt{{\cal M}(q')
{\cal M}(q)}}{2 \pi} \int d\vec{p}\,d\vec{P}\,\Psi_{f}^{*}(\vec{P},\vec{p}\,')
\,\hat{t}(\omega,\vec{q'},\vec{q},\vec{p}_{1})\,\Psi_{i}(\vec{P},\vec{p})\,,
\end{equation}
where the arguments of the nuclear wave functions are the Lovelace coordinates
\begin{equation}
\vec{p}=-\frac{\sqrt{3}}{2}\vec{p}_1\,,\qquad \vec{p}\,'=\vec{p}-\frac{\vec{Q}}
{\sqrt{3}}\,,\qquad \vec{P}=\frac{1}{2}(\vec{p}_{2}-\vec{p}_{3})\,\,,
\end{equation}
with the momentum transfer $\vec{Q}=\vec{q}-\vec{q}\,'$.

The nuclear wave functions $\Psi(\vec{P},\vec{p})$ are expanded in orbital
momentum, spin, and isospin space as
\begin{equation}
\Psi(\vec{P},\vec{p}) = \sum_{\alpha}\phi_{\alpha}(P,p) \mid (\tilde{L}l)
{\cal L},(\tilde{S}{\textstyle \frac{1}{2}}){\cal S},
{\textstyle \frac{1}{2}}M>\mid ({\tilde T}{\textstyle \frac{1}{2}})
{\textstyle \frac{1}{2}}\nu>\,,
\end{equation}
where $\phi_{\alpha}(P,p)$ are numerical solutions of the Fadeev equations.
To shorten the notation we introduced $\alpha={\{\tilde L}l{\cal L}{\tilde S}
{\cal S}{\tilde T\}}$, where ${\tilde L}$, ${\tilde S}$ and ${\tilde T}$
are the total angular momentum, spin, and isospin of the pair (2,3)
(${\tilde L}$ is associated with momentum $\vec{P}$), and $l$ and $\frac{1}{2}$
have an analogius meaning for the particle (1) ($l$ is associated with
momentum $\vec{p}$).

Nucleonic Fermi motion is treated as in Refs.\cite{35,36} where it was
found that the substitution
\begin{equation}
\vec{p}_{1} \rightarrow \vec{p}_{\rm eff.} = -\frac{\vec{q}}{A} -
\frac{A-1}{2A}(\vec{q} - \vec{q}\,')
\end{equation}
(factorization approximation) provides numerical results very close to the
exact ones. This approximation allows us to express the isospin
dependent potential
${\rm V}_{I}(\vec{q}\,',\vec{q})$ in the form
\begin{equation}
{\rm V}_{I}\!=\!4\pi W_{A}(-1)^{\frac{1}{2}-M_f}\!\!\sum_{STLJ}\!i^{L+S}
C_{I}^{T}A_{ST} M^{JT}_{SL}(Q)\!\left[K^{S}\!\times\!Y^{*L}(\hat{\vec{Q}})
\right]^{J}_{M_{J}}\!
\left(\;\frac{1}{2}\;\;\;\;J\;\;\;\;\frac{1}{2}
\atop{{\rm -}M_{f}\,M_{J}\,M_{i}}\right)\,,\;\;\;\;\;
\end{equation}
where $A_{ST}(\omega,\cos{\Theta^*})$ are the $\pi N$ scattering amplitudes
defined in eq.(4), $K^{0}=1$ and $K^{1}=[\hat{\vec{q}}_{i} \times
\hat{\vec{q}}_{f}]$\,,\, and $C_{I}^{T}$ and $W_{A}$ stand for the\\
isospin
\begin{equation}
C_{1/2}^{T}=(-1)^{T}\sqrt{\frac{T+1}{T+2}}\,,\qquad
C_{3/2}^{T}=\sqrt{\frac{1}{2(2T+1)}}
\end{equation}
and kinematical
\begin{equation}
W_{A}(\vec{q}\,',\vec{q}) = \sqrt{\frac{{\cal{M}}(q\,') {\cal{M}}(q)}
{ \mu (\vec{q}\,',\vec{p}_{1}\,') \mu (\vec{q},\vec{p}_{1})}}
\end{equation}
factors.
Using Fadeev wave functions from eq.(18) we obtained the following expression
for the nuclear form factor $M_{SL}^{JT}(Q)$,
\begin{mathletters}
\begin{eqnarray}
M_{SL}^{JT}(Q) & = & \frac{3}{\sqrt{4\pi}}i^{-L}\sum_{\alpha',\alpha}
O_{SL}^{JT}(\alpha',\alpha) \int d\vec{p}\left[Y^{l'}(\Omega_{\vec{p}\,'})
\times Y^{l}(\Omega_{\vec{p}}) \right]^{L}_{0} \nonumber\\
& & \times \int P^{2} dP \phi_{\alpha'}(P,p') \phi_{\alpha}(P,p)\,,
\end{eqnarray}
\begin{eqnarray}
O_{SL}^{JT}(\alpha',\alpha) &=& 8 \hat{\cal L}\hat{\cal L'}\hat{\cal S}
\hat{\cal S'}\hat{\tilde S}\hat{L}\hat{T}\delta_{\tilde{L}\tilde{L'}}
\delta_{\tilde{S}\tilde{S'}}\delta_{\tilde{T}\tilde{T'}}
(-1)^{l+l'+\tilde{L}-{\cal L+S}+\tilde{S}+\tilde{T}+S+T+3/2}
\nonumber\\ & & \times
\left\{ {\frac{1}{2}\,\,\,\;\frac{1}{2}\,\,\;S}\atop{{\cal S'}\;{\cal S}
\;{\tilde S}}\right\}
\left\{{{\cal L}\;{\cal L'}\;L}\atop{l'\,\,\,\;l\,\,\,\;{\tilde L}}
\right\}
\left\{{\frac{1}{2}\;\;{\tilde T}\;\;\frac{1}{2}}
\atop{\frac{1}{2}\;\;T\;\;\frac{1}{2}}
\right\}
\left\{ \begin{array}{ccc}
{\cal L} & {\cal S} & \frac{1}{2} \\
{\cal L'}& {\cal S'}& \frac{1}{2} \\
      L     &      S      &    J
\end{array} \right\}\,,
\end{eqnarray}
\end{mathletters}
where $\hat{a}=\sqrt{2a+1},\;$ and
$\left\{ \begin{array}{c}\cdots \\ \cdots \end{array} \right\}$ and
$\left\{ \begin{array}{c}\cdots \\ \cdots \\ \cdots \end{array} \right\}$
stand for the $6j$ and $9j$ symbols, respectively.

\subsection{The phenomenological model}

In eq.(23) parity and angular momentum selection rules determine
which nuclear form factors can contribute. In analogy with
electromagnetic form factors we refer to them as\\
$ S=0,\quad L=0,\,\ \quad J=0,\quad T=0$ or 1 -isoscalar or
isovector C0 form factor\\
$\quad S=1,\quad L=0,2\quad  J=1,\quad T=0$ or 1 -isoscalar or
isovector M1 form factor

In fact, assuming that mesonic exchange currents (MEC) contributions are
small in elastic electron scattering these form factors had been extracted
directly from the electromagnetic form factors of $^3$He and $^3$H\cite{4,37}.
It was believed that this assumption is more reliable for the matter
distribution described by C0 form factors. However, recent theoretical
investigations of the charge distributions of trinucleon systems show that
MEC contributions in C0 transitions at transfer momentum $Q^2>5 fm^{-2}$ are
important too\cite{38}. Therefore, the connection between matter distribution
and experimental charge form factors is not straightforward as it was
suggested in many previous calculations.

In this paper we try to estimate the upper limit of uncertainty, which was
present in early investigations by comparison of results obtained in
phenomenological models with microscopical calculations. In our
phenomenological model we express the C0 form factors by the charge form
factors of $^3$He (F$_{ch}^{^3\!{\rm He}}$)\cite{39} and
$^3$H (F$_{ch}^{^3\!{\rm H}}$)\cite{40}:
\begin{equation}
M_{00}^{0T}(Q)=\frac{1}{\sqrt{\pi}}\hat{T}[2{\rm F}_{ch}^{^3\!{\rm He}}(Q)
+(-1)^{T}{\rm F}_{ch}^{^3\!{\rm H}}(Q)]/{\rm f}_{ch}^{p}(Q)\,,
\end{equation}
where ${\rm f}_{ch}^{p}(Q)$ is the proton charge form factor.

For the M1 form factors we will keep the microscopical description via the
Fadeev wave function, because here the phenomenological approach is clearly
not applicable. First, there are large MEC contributions in the M1 electron
scattering form factor, and secondly, it is not possible to unambiguously
separate the contributions from spin and convection currents. Therefore,
we will use the Fadeev wave functions described in the previous subsection
for the J=1 transition densities.

\subsection{The simple S-shell model}

Finally, we employ a very simple model where  the nucleons
are in the $0s_{1/2}$ shell of a harmonic oscillator potential
and, furthermore, pion rescattering
and Coulomb contributions are neglected,
 e.g. ${\rm F}_{\pi' \pi} = {\rm V}_{\pi' \pi}^{(1)}$
(plane wave approximation). The last assumption allows expressing
${\cal F}_{I}$ and ${\cal G}_{I}$, introduced in the previous section,
directly via the non spin-flip ($A_{00}, A_{01}$) and spin flip
($A_{10}, A_{11}$) components of the $\pi N$ elementary amplitude of eq.(4).

Evaluating eq.(20) in the frame with the initial pion momentum $\vec q$
along the z-axis and the vector $[\vec{q} \times \vec{q}\,']$ along the
y-axis (as in eqs.(12-15)) the plane wave  expressions for
${\cal F}_{I}$ and ${\cal G}_{I}$ can be written as
\begin{mathletters}
\begin{equation}
{\cal F}_{I}=\sqrt{2\pi}W_{A} \sum_{T} C_{I}^{T}\,
A_{0T}(\omega,\cos{\Theta^*})\,M_{00}^{0T}(Q)
\end{equation}
\begin{equation}
{\cal G}_{I}=\sqrt{\frac{2\pi}{3}}W_{A}W_{B} \sum_{T} C_{I}^{T}\,
A_{1T}(\omega,\cos{\Theta^*})\,[\,M_{10}^{1T}(Q) -
\frac{1}{\sqrt{2}}M_{12}^{1T}(Q)\,]\,,
\end{equation}
\end{mathletters}
where $W_{B}$ is an additional kinematical factor arising due to the Lorentz
transformation of the vector $[\hat{\vec{q}}_{i} \times \hat{\vec{q}}_{f}]$
in the elementary amplitude from the $\pi N$ c.m. to the $\pi A$ c.m. system.
Note that the expressions (25) are valid not only for $^3$He or $^3$H but
for all nuclei with spin and isospin one half ($^{13}$C, $^{15}$N, \ldots)

In the framework of the S-shell model the nuclear form factors $M_{SL}^{JT}(Q)$
are simply given by
\begin{equation}
M_{SL}^{JT}(Q)=\frac{1}{\sqrt{\pi}}\,{\hat T}{\hat S}\,R_{S}(Q)\,\delta_{L,0}
\,\Psi_{JT}\,,
\end{equation}
where $R_{S}(Q)=\exp{(-b^{2}Q^{2}/6})$ with $b$=1.65 fm as the point radius of
$^3$He, and the
coefficients are $\Psi_{00}$=3 and $\Psi_{01}=\Psi_{10}=-\Psi_{11}=1$.
In this case we can express ${\cal F}$ and ${\cal G}$ for the
reactions (11) via the analogous amplitudes $f$ and $g$ for $\pi N$
scattering:\\
for elastic $\pi^{\pm}-^3$He or $\pi^{\mp}-^3$H scattering
\begin{mathletters}
\begin{equation}
{\cal F}=(2f_{\pi^\pm{\rm p}}+f_{\pi^\pm{\rm n}})R_{S}(Q)W_{A}=
(3A_{00}\pm A_{01})R_{S}(Q)W_{A}
\end{equation}
\begin{equation}
{\cal G}=g_{\pi^\pm{\rm n}}R_{S}(Q)W_{A}W_{B}=
(A_{10}\mp A_{11})R_{S}(Q)W_{A}W_{B}
\end{equation}
for the $^3$He$(\pi^-,\pi^0)^3$H or $^3$H$(\pi^+,\pi^0)^3$He reactions
\end{mathletters}
\begin{mathletters}
\begin{equation}
{\cal F}=\frac{1}{\sqrt{2}}(f_{\pi^+{\rm p}}-f_{\pi^+{\rm n}})R_{S}(Q)W_{A}=
\sqrt{2}A_{01}R_{S}(Q)W_{A}
\end{equation}
\begin{equation}
{\cal G} =-\frac{1}{\sqrt{2}}(g_{\pi^+{\rm p}}-g_{\pi^+{\rm n}})R_{S}(Q)W_{A}
W_{B}=-\sqrt{2}A_{11}R_{S}(Q)W_{A}W_{B}\,.
\end{equation}
\end{mathletters}
The above expressions show
that in the S-shell model all information
about nuclear structure is contained in the form factor $R_{S}(Q)$.
It divides out in the expression for the asymmetry (eq.(15)); thus, $A_{y}$
is given in terms of the free $\pi N$ scattering amplitudes only.

\section{ RESULTS AND DISCUSSION}

\subsection{Pion scattering on unpolarized  targets}

We begin our discussion with the analysis of some  main features of
the pion-nuclear interaction in the energy region of $T_{\pi}$=100--300 MeV.
The corresponding results of our calculations are depicted in Fig.1.

One of the important properties of the $\pi N$ interaction in this energy
region is the importance of the $\Delta$-isobar excitation, especially
around $T_{\pi}$=200 MeV and, consequently, the large contribution from pion
$p$-waves. This feature is reflected in the coherent scattering process which
is proportional to $A$ (nuclear mass number) and described by the scalar part
($A_{00}$) of the $\pi N$ amplitude. Since the $p$-wave part of this
amplitude has a $\cos{\Theta}$ dependence ( see Appendix) the differential
cross section experiences a minimum around $\Theta$=90$^{\circ}$. The position
of this minimum is in fact shifted due to the Lorentz transformation of the
pion angle from the $\pi N$ to the $\pi-^3$He c.m. frame and due to the
$s$-, and $p$-waves interference. With increasing pion energy the minimum
disappears because the contributions of the other multipoles become larger.

The spin-flip transitions due to the amplitudes $A_{10}$
 and $A_{11}$ are proportional to $\sin{\Theta}$ which in the
$\Delta$-resonance region fills in the minimum. Note that for
 $\pi^-$ scattering the minimum is filled in more than for $\pi^+$ scattering.
This  can easily be understood in the framework of the S-shell
model where the spins  of the two protons are coupled to zero. Therefore,
due to the Pauli principle the spin-flip transition can be realized
only through the neutron distribution in $^3$He. However, it is well known
that in the $\Delta$-resonance region the $\pi^- n$ interaction is about
10 times stronger than the $\pi^+ n$ one. Hence the strength of the spin-flip
transition in $\pi^{-}\,^3$He scattering is about one order of magnitude larger
than in $\pi^{+}\,^3$He scattering.

In Fig.1 we compare the results obtained with correlated three-body Fadeev
wave functions with calculations performed in the simple S-shell model.
At pion energies up to $T_{\pi}$=300 MeV  the momentum transfer to the nucleus
approaches $Q=3.6 fm^{-1}$ which is not yet sufficient
to differentiate between the phenomenological and the full three-body model.
At $T_{\pi}$=100 MeV even the simple S-shell model agrees well with our full
calculation. Thus, at this low energy plane wave calculations with S-shell
harmonic oscillator wave functions are sufficient not only for a qualitative
but also for a quantitative discussion of the $\pi^{+}\,^3$He interaction.
The agreement with experimental data from Refs.\cite{41,42} in all three
models is very good.

Moving into the region of the $\Delta$-isobar excitation the contribution of
pion rescattering becomes larger and reaches its maximum at
$T_{\pi}$=180--200 MeV. At backward angles the S-shell model calculations
(with rescattering) deviate significantly from our full calculations in the
three-body model indicating that S-shell harmonic oscillator wave functions
have become insufficient to describe the nuclear wave function.
At $T_{\pi}$=295 MeV the full calculations fail to reproduce the experimental
data from Ref.\cite{41} at large angles both in magnitude and in the shape of
the differential cross section. The measurements indicate an additional dip
around $\Theta$=120$^{\circ}$ while our calculations yield smooth predictions
for $d\sigma/d\Omega$. Furthermore, around $\Theta$=90$^{\circ}$  our
computation underestimates the data in $\pi^-$ scattering but overestimates
them in $\pi^+$ scattering. Measurements of polarization observables may be
useful in this kinematic region since asymmetries tend to be large where
angular distributions have minima and could thus be sensitive to subtleties
in the reaction mechanism.

We developed a similar approach for the  pion-deuteron interaction
using a two-body wave function generated by the Paris potential\cite{43}.
In Fig.2 we present our results for $\pi^+\,d$ elastic scattering at
$T_{\pi}$=180--300 MeV. Here the agreement with experimental data is
excellent even at backward angles. Note that the main reason of that may be
connected with the dominance of $\Delta$-resonance contribution. In this
case as have been shown in Ref.\cite{45} two-body approach is a good
approximation for the more elaborate three-body Fadeev calculations.

However at lower energies ($T_{\pi}< 50$ MeV) our model  fails to reproduce
the experimental data for the elastic $\pi d$ scattering. This is due to the
well known problem of the description of the $S$-wave pion-nucleon interaction
in nuclei with zero isospin. In such nuclei the contribution from the
large isovector $\pi N$ scattering lengths are canceled. Therefore the role
of higher order effects in the pion-nuclear potentials become important.
But in the case of $^3$He at low energies the scattering length is entirely
formed by the isovector part of pion S-wave scattering on the additional
proton.  Therefore in this case our model with first order optical
potential works much better\cite{19}.

Since $\pi\,d$ scattering in contrast
to $\pi\,^3$He is realized only via isoscalar transitions we assume the
discrepancy at backward angles in $\pi\,^3$He elastic scattering at
$T_{\pi}=295$ MeV to be due to isovector second order effects. We will
continue discussing this issue in subsection C where we consider the pion
charge exchange reaction which is described solely by the isovector part of
the pion-nuclear interaction.

Fig.3 presents our results for higher pion energies, $T_{\pi}$=350--500 MeV;
experiments in this energy domain have been proposed at LAMPF. At these
energies the momentum transfer to $^3$He for backward pion angles approaches
$Q=5 fm^{-1}$. These Q-values are large enough for differences between our
three-body and phenomenological model to become visible.
Therefore, applying phenomenological models in this region becomes
questionable since at these momentum transfers meson exchange currents
give important contributions to the charge form factors.

As mentioned before the $p$-wave dominance of the pion-nuclear interaction
decreases with increasing of pion energy but it does not disappear entirely.
There is still a noticeable deviation from the exponential fall of
$d\sigma/d\Omega$ around $\Theta$=90$^{\circ}$  in terms of a broad bump.
At larger angles the angular distributions obtained with both models
go through a minimum that comes from the C0 nuclear form factor. However,
even though the phenomenological model predicts this dip at smaller angles
than the three-body model the $\pi^+$ scattering data at $T_{\pi}$=350 MeV
suggest a minimum at even smaller pion angles. No conclusions are possible for
$\pi^-$ scattering or higher pion energies since the experimental information
is insufficient.

\subsection{Asymmetry and superratio.}

   In Fig.4, we present our results for the asymmetry $A_{y}$ at $T_{\pi}$
=100, 300 and 500 MeV both for $\pi^+$ and $\pi^-$ scattering.
In $\pi^+$ scattering $A_{y}$ approaches its maximum value of almost +1
at $T_{\pi}$=100 MeV around $\Theta$=90$^{\circ}$ (Fig. 4a).This is due again
to the $p$-wave dominance of the $\pi N$ interaction which also causes the
differential cross section to go through a minimum near 90$^{\circ}$. This
large asymmetry is not obvious since the analyzing powers of the elementary
$\pi N$ reactions are quite small. In fact, $ -0.2<A_y<0.08$ in $\pi^+ n$
scattering and $0<A_y<0.5$ in $\pi^+ p$ scattering. This indicates that the
view of $^3$He as a neutron target in the case of elastic pion scattering
would lead to the wrong conclusions. While the spin-flip amplitude is similar
to that of a neutron target, the non-spin-flip term is quite different.
A detailed analysis of the asymmetry in this region has been performed in
a previous paper\cite{20}. Up to $T_{\pi}$=180 MeV the results of the simple
S-shell model are very similar to the full calculations in the three-body
model.
Therefore, we conclude that in this region the asymmetry contains very
little nuclear structure information. Furthermore, the pion rescattering
effects are minimal at these energies.

 However, this situation changes drastically at higher energies.
In the full calculation $A_y$ goes through zero around $T_{\pi}$=200 MeV and
becomes large and negative in the 260--300 MeV region. In contrast, the
asymmetry in the simple model remains positive at higher energies.
As discussed in ref.\cite{20} pion rescattering
is mainly responsible for this effect.

Our results at $T_{\pi}$=100 MeV have been confirmed by first experimental
measurements from TRIUMF\cite{42}. As can be seen from Fig.4a, except for
the fact that the maximum of the calculated asymmetry $A_y$ appears to lie at
slightly smaller $\Theta$, the agreement with the data is very good.

In case of  $\pi^-$ scattering at $T_{\pi}$=100 MeV (Fig.4b) the difference
 between the simple
model and the full calculation is larger than for the $\pi^+$ case. This is
caused mainly by the larger influence of pion rescattering. At $\Theta$
=90$^{\circ}$  the differential cross section shows no minimum in contrast to
$\pi^+$ scattering. Therefore, the absolute value of $A_y$ is smaller and it
has a less pronounced structure in the angular distribution.

In the high energy region around $T_{\pi}$= 500 MeV the contribution of the
$D_{13}$(1525) resonance to the spin-flip part of the elementary amplitude
becomes important and $A_{10}$ and $A_{11}$ can approximately be written as
$A_{11}=-A_{10}\approx D_{13} \cos{\Theta}$ (see Appendix). Thus, for $\pi^+$
scattering the spin-flip amplitude  $G$ is proportional to
($A_{10}-A_{11}$) while in case of $\pi^-$ scattering $G$ is proportional to
($A_{10}+A_{11}$) which vanishes. Therefore, the asymmetry for $\pi^-$
scattering at $T_{\pi}$=500 MeV is zero almost everywhere in the simple S-shell
model. However, this exact cancellation is destroyed in the forward
direction by pion rescattering and at backward angles by the D-state
components of the $^3$He wave function.

In the case of $\pi^+$ scattering at $T_{\pi}$=500 MeV, where $G\approx D_{13}
\cos{\Theta}$, the asymmetry in the forward direction is large and
it depends neither on the nuclear model nor on pion rescattering.
Note that similar results for the asymmetry at $T_{\pi}$=500 MeV have been
obtained by Chakravarti et al.\cite{47}. Thus, in this kinematical region
the asymmetry can be described directly via the elementary amplitude
in accordance with eq.(27). Measurements in this region could extract
information on the $D_{13}$ resonance in the nuclear medium.

Concluding our analysis of the elastic channel in the $\pi\,^3$He interaction
we briefly consider the so-called "superratio"
\begin{equation}
 R=\frac{d\sigma(\pi^{+}\,^3{\rm H})\,d\sigma(\pi^{-}\,^3{\rm H})}
       {d\sigma(\pi^{+}\,^3{\rm He})\,d\sigma(\pi^{-}\,^3{\rm He})}
\end{equation}
discussed in detail in Ref.\cite{11}. The main point of interest
is related to the attempt to find charge symmetry breaking in the
strong interaction. It is expected that $R$ is insensitive to the
model uncertainties in the pion-nuclear interaction and
to the Coulomb interaction. Therefore, in accordance with the relations
(11) this ratio would have to be 1 at all angles and energies. But measurements
\cite{48,49,50} obtained a value $R > 1$ along with angular and energy
dependence. The generally accepted explanation \cite{11,51} for this
significant charge symmetry breaking effect assumes that the 2\% difference
in the $^3$He and $^3$H radius caused by the Coulomb repulsion between the two
protons in $^3$He is responsible. In Fig. 5 we compare our full model with the
results of our calculations in the S-shell model using

\begin{tabular}{lll}
1) & $b=1.65 fm$ & both for $^3$H and $^3$He (dashed curves)\\
2) & $b=1.65 fm$ & for $^3$H and $b=1.68 fm$ for $^3$He
(dash-dotted curves).
\end{tabular}

Using different harmonic oscillator parameters for $^3$H and $^3$He allows us
to qualitatively explain the measured values for $R$, except for the region
where the differential cross sections have the minimum.
The deviation from experiment at larger angles is clearly due to the
fact that harmonic oscillator wave functions at backward angles
are inappropriate at these momentum transfers (which follows
from our analysis of differential cross sections in the previous subsection).
In this region we certainly require a realistic three-body wave function
of $^3$He that includes the additional Coulomb interaction between the two
protons\cite{52}, like the one used by Gibbs and Gibson\cite{11}.
Our three-body model gives $R\approx 1$ at all angles ( except small angles).

In Fig.6 we show the differential cross sections at $T_{\pi}$=180 MeV for all
of the 4 reactions entering the superratio calculated with isospin symmetric
Fadeev wave functions. At backward angles preliminary data of Briscoe
et al.\cite{50} show deviation from our calculations which we did not find
for the $T_{\pi}$=200 MeV data in Fig.1. As in our discussion about the
discrepancies  with the $T_{\pi}$=295 MeV data, we find in accordance to
Ref.\cite{14} the need of second order effects. To illustrate this we apply
the $\rho^2$-term of the pion-nucleus interaction from Ref.\cite{11}.
In this way we obtain a better agreement with the data, however, in the
case $\pi^-\,^3$H and $\pi^+\,^3$He a large deviation remains at
$\Theta>$150$^{\circ}$.

\subsection{Pion single charge exchange}

In the previous subsections we have demonstrated that our formalism
generally gives a good description of the elastic scattering data both for
$\pi^+$ and $\pi^-$ at pion angles 0$^{\circ}$$<\Theta<120$$^{\circ}$.
Assuming charge
symmetry it follows that the isovector part of the pion-nuclear interaction
which is responsible for the pion single charge exchange (SCE) reaction is
accurately described in this kinematical region. Therefore we should not
encounter serious difficulties in the description of the SCE process.

The results of our coupled channels calculations for
$^3$H($\pi^+,\pi^0)^3$He at $T_{\pi}$=130 MeV and for $^3$He($\pi^-,\pi^0)^3$H
at $T_{\pi}$=200 MeV are presented in Fig.7. Note that the rescattering
contributions in SCE are larger than in the elastic channel.
This is due mainly to the incoherent nature of the SCE process. Therefore,
the influence of the elastic channel (proportional to $A$) on the pion
rescattering term becomes enhanced. Experimental data at backward angles
obtained in Ref.\cite{53} by detecting the recoiling $^3$He are in excellent
agreement with our calculations performed in the framework of the three-body
model. Similar results have been obtained by Landau \cite{4,6} with
phenomenological nuclear form factors.

The SCE reaction on $^3$He at $T_{\pi}$=200 MeV is at the present time
the only example where experimental data at forward angles have become
available\cite{54} by directly detecting the $\pi^0$. The comparison
with our full calculation shows a very good agreement  up to
$\Theta$=60--70$^{\circ}$. However, at larger angles there is a small
deviation which, as we will see below, grows with increasing pion energy.

In Fig.7 we also show the contribution from the spin-flip and non-spin-flip
part of the elementary $\pi N$ scattering amplitude separately. As in the case
of elastic scattering the non-spin-flip transition experiences a minimum
around $\Theta$=90$^{\circ}$ due again to the $p$-wave nature of the $\pi N$
interaction. The spin-flip contribution which has a $\sin{\Theta}$ angular
dependence fills in this minimum. Note that for the SCE reaction the relative
strength of the spin-flip transition is larger compared to elastic scattering.
This can be attributed to the incoherent nature of both the spin-flip and the
non-spin-flip mechanism of the SCE process.

Fig.7 demonstrates that there is interference between spin-flip
and non-spin-flip transitions due entirely to rescattering effects.
For example, the non-spin-flip transition in PWIA can be realized only
through the non-spin-flip part of the $\pi N$ amplitude; in this case no
interference is present. However, if pion rescattering is taken into account
in a coupled channels framework the non-spin-flip transitions can be realized
through double spin-flip transitions as well.

We now proceed to consider the high-energy region. The results depicted
in Fig.8 show dramatic discrepancies between theory and experiment.
Our coupled-channels calculations with three-body wave functions
underestimate the data at backward angles about two orders of magnitude at
$T_{\pi}$=285 MeV and about one order of magnitude at higher energies.

At present there is no explanation for this disagreement except some
calculations in a Glauber multiple-scattering formalism
performed in Refs.\cite{12,13}. However, we feel that this approach is
inappropriate for an analysis in the large scattering-angle region.

As discussed in subsection B, one of the reasons for such a large
discrepancy could be due to second-order contributions in the isovector
part of the pion-nuclear interaction. A microscopic analysis of corresponding
effects in $\pi\,^3$He scattering has been performed by Wakamatsu\cite{14}.
However, the contributions he found are not large enough to explain
discrepancies of two orders of magnitude. One can remove this discrepancy
by artificially enhancing the isovector second-order interaction whose
influence on the SCE reactions is much stronger  than on  elastic ones.
To check this assumption qualitatively we followed the prescription given
in Refs.\cite{8,14} and introduced a second-order potential of the form
\begin{equation}
{\rm V}_{\pi' \pi}^{(2)}(\vec{q}\,',\vec{q})\sim (A-1)^2\,
\left( B_1\,+\,C_1\,\vec{q\,'}\cdot\vec{q}\right)\,
exp(-\frac{bQ^2}{8})\,\vec{\tau}\cdot\vec{t}
\end{equation}
with real parameters $B_1$ and $C_1$. Treating $B_1$ and $C_1$ as free
parameters we extract $B_1=0.086/m_{\pi}^4$ and $C_1=-0.058/m_{\pi}^6$
from a fit to the SCE data. These values are of similar magnitude as the
parameters $B_0$ and $C_0$ extracted from an isoscalar second-order potential
in Ref.\cite{8}. In principle, $B_1$ and $C_1$ would have to be complex.
However, a microscopic derivation of $V^{(2)}$ would in some way involve the
square of the  elementary $\pi N$ t-matrix and since the imaginary part
dominates this amplitude around $T_{\pi}$=300 MeV the main contribution to
$B_1$ and $C_1$ should be real. The results of our calculation including
$V^{(2)}$ are shown in Fig.9. We confirm that the contribution of this
potential in the charge-exchange channel is in fact much larger than
in the elastic one. At the same time the second-order effects
in elastic scattering at backward angles are of the same order as the
discrepancy with experimental data. Thus, the longstanding problem
in the description of the SCE reaction and elastic scattering at $T_{\pi}$=300
MeV at backward angles might have the same origin. Again, an isoscalar
second-order potential \cite{8} has to be included in the elastic scattering
reactions before final conclusions can be drawn. To clarify the situation
additional theoretical studies and experimental measurements
in this region are required.

Note that the necessity of a second order potential follows also from
three-body unitarity which requires to extend the nuclear model space by
including the breakup channels. Our results show that the coupling with
this channels is probably very important at large momentum transfer,
in particular for SCE reactions at high energies and backward angles.

In Fig.10 we present results for the asymmetry $A_y$ in the SCE
reaction on $^3$He at $T_{\pi}$=100, 300 and 500 MeV. As in the case of elastic
scattering at $T_{\pi}$=100 MeV the simple S-shell model approximately
reproduces the results of the full calculation. The small difference is caused
mainly by pion rescattering. Note that if in eq.(28)
for the simple model we neglect the small kinematic correction from the
factor $W_B$ we obtain a simple relation between $A_y$ for $^3$He and
the elementary asymmetry:
\begin{equation}
                 A_{y}(^3{\rm He})\approx - A_{y}(p)
\end{equation}
The minus sign in this relation results from the opposite sign between
spin-flip and non-spin-flip nuclear matrix elements  (see eq.(26)) in
$^3$He($\pi^-,\pi^0)^3$H compared to the free process $p(\pi^-,\pi^0)n$.

Moving into the $\Delta$ resonance region the pion rescattering
contribution becomes enhanced changing the sign of the asymmetry around
$T_{\pi}$=220 MeV. However, increasing the pion energy decreases  the role of
pion rescattering. Therefore, at $T_{\pi}$=500 MeV in the forward direction the
simple S-shell model results are again close to the full calculations.
In this region the simple relation (31) is again fulfilled as in the case of
$T_{\pi}$=100 MeV. However, while at lower energies the role of the $P_{33}$
wave was dominating $A_y$, at 500 MeV it is the $D_{13}$ resonance which has
become important.

\subsection{ Pion induced eta production}

Eta production by pions is another important $\pi N$ inelastic
channel. Interest in the physics with eta mesons has grown significantly
in recent years, experiments using hadronic probes to produce $\eta$ mesons
have been performed at LAMPF, Brookhaven and Saclay. On the other hand,
electron accelerators such as BATES, ELSA and the Mainz Microtron
open the possibility to produce eta mesons with electrons or real photons.

At the present time little is known about the nature of the eta-nucleus
interaction.
For the elementary $\pi N\rightarrow\eta N$ process the experimental data is
much less complete and accurate in contrast to $\pi N$-scattering data.
There are also only few theoretical investigations of this reaction
\cite{22,23,24} based on the coupled channel isobar model for the $\pi N,\,
\, \eta N$ and $\pi\pi N$ systems without background. In accordance with a
recent analysis by Benmerrouche and Mukhopadhyay\cite{57}, where eta
photoproduction has been studied, the role of background could be important.
However, the corresponding contribution strongly depends on the value of
the not well-defined $\eta NN$ coupling constant
$g_{\eta}\,\,(0.6 \leq g_{\eta}^2/4\pi\leq 6.3)$. In such a situation the
investigations of pion induced eta production on lightest nuclei could give
additional information about the elementary amplitude. But before that we
have to be sure that all other ingredients connected with the reaction
mechanism and nuclear structure input  are well under control.

     Below we will concentrate mainly on the study of the initial and final
state interaction and nuclear structure effects. The elementary processes
with eta mesons will be described in the framework of the coupled channels
isobar model of Ref.\cite{24}  with parameters extracted from available data
( see also Appendix).

The amplitudes for the nuclear processes have
been obtained by solving the system of equations similar to eq.(10) but
extended to include the $\eta$ channels. In this case ( omitting the index
$L_{\pi}$) eq.(10) can be rewritten the following way
\begin{equation}
F^{j,I}_{n',n}(q',q)=V^{j,I}_{n',n}(q',q)-\frac{a}{\pi}\sum_{n''}\int
\frac{q''^2}{{\cal{M}}(q'')}\frac{V^{j,I}_{n',n''}(q',q'')
F^{j,I}_{n'',n}(q'',q)}{E(q) - E(q'') + i\epsilon} dq''
\end{equation}
where $n=\pi,\,\eta$ with $\pi$ and $\eta$ labelling the $\pi A$ and
$\eta A$ channels, respectively.

Since the isospin of the eta is zero, only contributions
to the channel with total isospin 1/2 are possible.
 The definition of the corresponding isospin dependent
amplitudes F$_I$ or V$_I$ is the same as in eqs.(8) and (20).
However,the isospin factors $C_{I}^{T}$ in eq.(20) have to be changed to
\begin{mathletters}
\begin{equation}
C_{I}^{T}= \frac{1}{\sqrt{2}}\delta_{I,\frac{1}{2}} \delta_{T,0}
\end{equation}
for $(\eta,\eta)$ scattering and for the $(\pi,\eta)$ reaction to
\begin{equation}
C_{I}^{T}=-\frac{1}{\sqrt{2}}\delta_{I,\frac{1}{2}} \delta_{T,1}\,.
\end{equation}
\end{mathletters}

Results of our calculations for the differential cross section at
$T_{\pi}$=555 MeV are shown in Fig.11. First we point out that we
achieve a good description of recent measurements\cite{25} in the forward
direction. Our agreement with the data is in contrast to the DWIA
calculations in Ref.\cite{25} which underestimate the data by a factor of
about 3.
The corrections from two-step processes, such as $(\pi^-,\pi^0)(\pi^0,\eta)$,
are not very large (about 10\%). The main effect of the pion- and eta-nuclear
interaction in the initial and final state is to fill in the diffraction
minimum in the differential cross section.

The situation changes dramatically in the backward direction.
Here our full calculation with three-body wave functions underestimates
the experimental measurements by a factor of about 50. Employing
phenomenological nuclear form factors for the J=0 transition densities
reduces the disagreement with the data to a factor of 2-3, similar to the
findings of Ref.\cite{25}. However, the application of the phenomenological
approach, as mentioned before, is questionable because we are again in a high
momentum transfer region. The origin of this discrepancy may be similar to
the one encountered in the pion SCE reaction.

In contrast to the differential cross section the asymmetry $A_y$ is less
sensitive to the details of nuclear structure. Therefore, in the forward
direction, as in the pion SCE reaction,  we can approximately write
\begin{equation}
 A_{y}^{\pi,\eta}(^3{\rm He}) \approx -A_{y}^{\pi,\eta}(p)
\end{equation}

Fig.12 presents the dependence of the differential cross section at
$\Theta_{\eta}$=0$^{\circ}$ on the eta momentum in the $\eta ^3$He c.m.
system. These data can be reproduced only for $q_{\eta} >150$ MeV/c which
corresponds to $T_{\pi}$$>$490 MeV, while at lower eta momenta our
calculations (similar to Ref.\cite{25}) significantly underestimate the
observed cross section regardless which nuclear model is used.

As pointed out in Ref.\cite{25} this is a region where the $(\pi,\eta)$
reaction proceeds  below the free eta production threshold. Here the
corrections to the impulse approximation as well as different off-shell
behaviors of the elementary amplitude could be important. These effects may
lead to significant enhancements of the cross section similar to those
observed, for example, in $\pi^0$- photoproduction at threshold on very light
nuclei.\cite{58}

\section{CONCLUSION}

In this paper we have studied the interaction of pions with $^3$He and $^3$H
in a coupled channels, multiple-scattering approach carried out in momentum
space. Our investigation covered a wide energy region; from $T_{\pi}$=
100 MeV - well below the $\Delta$ resonance - beyond the $\Delta$ region into
the domain of the $D_{13}$(1525) resonance and up to the $\eta$ production
threshold. Correlated three-body Fadeev wave functions were employed to
describe the trinucleon ground state. Phase-shift parameterizations were used
for the elementary $\pi N$ amplitudes, along with a separable potential for the
off-shell extrapolation. The eta-production channel, $\pi N \rightarrow
\eta N$, was described in an isobar model that includes the $S_{11}$(1535),
$P_{11}$(1440) and $D_{13}$(1525) resonances as S-, P-, and D-wave interactions
and reproduces available low energy $\pi^- p \rightarrow \eta n$ cross
section data.

The  $\pi^+$ and $\pi^-$ elastic scattering on $^3$He
were well reproduced in our model in almost the entire kinematical region
considered here. Only for pion kinetic energies above 180 MeV and backward
pion angles - a region with few experimental data - hints of an
inadequate description appear. In this kinematical region phenomenological
$C0$ nuclear form factors extracted from charge form factors should not be
used any more because MEC contributions have become substantial.
The asymmetries for $\pi^+\,^3$He
elastic scattering are large in contrast to asymmetries measured on p-shell
nuclei. Noteworthy is the change of sign in $A_y$ from +1 to -1 when one
moves into the $\Delta$-resonance region; this effect - caused by pion
multiple scattering - should be verified experimentally. At $T_{\pi}$=500 MeV
and in the forward direction the asymmetry is entirely determined by the
$D_{13}$ resonance contribution.

In contrast to pion elastic scattering the pion single charge exchange
calculations agree with the data only up to $T_{\pi}$=200 MeV; at higher energy
the theoretical description dramatically fails to explain the measurements
by underestimating them up to two orders of magnitude. Since the $\pi N$
amplitudes and the nuclear wave function is presumably well known this
discrepancy may be an indication for two- and three-body processes that go
beyond the impulse approximation. We note that a similar phenomenon has been
observed in the photoproduction process $^3$He($\gamma,\pi^+$)$^3$H at large
momentum transfers. Since, on the other hand, $\pi^+\,d$ elastic scattering
data can be reproduced very well in our model we have introduced a
phenomenological isovector second-order potential and adjusted the parameters
to reproduce the SCE data. This is very much connected to the problems we
have found for back-angle elastic scattering at $T_{\pi}$=180 MeV. Further
investigations of second-order potential has to be done.
Clearly, it would be desirable to derive such
a potential microscopically. Again, the asymmetry of the process
$^3$He($\pi^-,\pi^0$)$^3$H is predicted to be large.

Finally, we discussed pion induced eta production, $^3$He($\pi^-,\eta$)$^3$H,
in the framework of our coupled-channels model. At $T_{\pi}$$<$ 490 MeV which
is a region below the free production threshold our results for forward
eta production significantly underestimate the data. We found good agreement
with the few available data at small momentum transfer and $T_{\pi}$$>$ 560 MeV
but large deviations in the backward direction with large Q. We believe that
the same mechanism in both incoherent processes is responsible for this puzzle.
Future  theoretical studies should reveal if these discrepancies present a
clear indication of a breakdown in the impulse approximation. Very recently
a paper by L. C. Liu has been published which confirms our conclusion on
the  importance of two-nucleon effects in nuclear eta production\cite{59}.

\acknowledgments

We want to thank W. J. Briscoe and S. K. Matthews for discussions and for
providing us with their preliminary data and R. Mach for discussions on the
pion-nuclear interaction.

This work was supported by the Deutsche Forschunsgemeinschaft
(SFB201) and the Natural Science and Engineering Research Council of Canada
(NSERC).

\unletteredappendix{}

In this appendix we define our partial wave amplitudes for
$\pi N$ scattering, eta production and $\eta N$ scattering
which in the past have been given by different authors using various
conventions.

As in eq.(4) the elementary amplitudes for all three
processes  are defined as:
\begin{equation}
\hat{f}_{m,n}=A_{00}+A_{01}\,\vec{t}_{m,n}\cdot\vec{\tau}+i\vec{\sigma}\cdot
\,[\hat{\vec{q}}_{i}\times\hat{\vec{q}}_{f}]\,
(A_{10}+A_{11}\,\vec{t}_{m,n}\cdot\vec{\tau})\,,
\end{equation}
where $n$ and $m$ label the $\pi$ or $\eta$ mesons. Here
we introduce the auxiliary matrix $\vec{t}_{m,n}$. The cyclic components
of this matrix are:

1)for $\pi N$ scattering  $(\vec{t}_{\pi,\pi})_{\lambda}=(\vec{t})_{\lambda}$
is the standard pion isospin operator;

2) for the $\pi N \leftrightarrow \eta N$
reaction $(\vec{t}_{\pi,\eta})_{\lambda}=(-1)^{\lambda}$ and

3) for $\eta N$-scattering $(\vec{t}_{\eta,\eta})_{\lambda}=0$.

The differential cross section for the
elementary processes in the meson-nucleon c.m.
system  is given  as
\begin{equation}
\frac{d\sigma_{m,n}}{d\Omega}=\frac{q_f}{2\,q_i}\,Tr\,\left[\,f_{m,n}^{+}
\,f_{m,n}\right].
\end{equation}

The expansion of the $A_{ST}$ amplitudes in partial amplitudes with orbital
angular momentum $l$ and total angular momentum $j=l\pm \frac{1}{2}$ is
identical for all three processes:\\
for non spin-flip ($S=0$) amplitudes
\begin{equation}
A_{0T}=\sum_{l}[\,(l+1)\,f_{lT}^{(+)}(w)+l\,f_{lT}^{(-)}(w)\,]\,
P_{l}(\cos{\theta})
\end{equation}
and for spin-flip amplitudes
\begin{equation}
A_{1T}=\sum_{l}[\,f_{lT}^{(+)}(w)-f_{lT}^{(-)}(w)\,]\,P'_{l}(\cos{\theta})\,.
\end{equation}

The main difference arising in the formalism for
$\pi N$ scattering, $\eta N$ scattering and $\pi N \leftrightarrow \eta N$
lies in the expansion of the partial amplitudes
$f_{lT}^{(\pm)}$ in terms of contributions with total isospin
$I=\frac{1}{2}\,\,(f_{l\pm}^{1/2})$ and $I=\frac{3}{2}\,\,(f_{l\pm}^{3/2})$;
where $\pm$ corresponds to $j=l\pm\frac{1}{2}$. Therefore, below these
reactions  are presented separately.

\begin{center}
{\bf 1. $\pi N$ scattering}
\end{center}
\begin{equation}
f_{l0}^{(\pm)}=\frac{1}{3}\left(2f_{l\pm}^{3/2}+f_{l\pm}^{1/2}\right)
\qquad{\rm{and}}\qquad
f_{l1}^{(\pm)}=\frac{1}{3}\left(f_{l\pm}^{3/2}-f_{l\pm}^{1/2}\right)\,.
\end{equation}

Keeping only the contributions from $S,\,P$ and
$D$ waves in the expansions (A3-A4) and using standard notation,
$f_{lj}^{I}\equiv l_{2I,2j}$, for the $A_{ST}$ amplitudes we arrive at
\begin{eqnarray}
3A_{00}&=& 2\,S_{31}+S_{11}+(4\,P_{33}+2\,P_{13}+2\,P_{31}+P_{11})\cos\theta+
\nonumber \\ & &
(3\,D_{15}+2\,D_{13})P_{2}(\cos\theta)\,,
\end{eqnarray}
\begin{eqnarray}
3A_{01}&=&S_{31}-S_{11}+(2\,P_{33}-2\,P_{13}+P_{31}-P_{11})\cos\theta-
\nonumber \\ & &
(3\,D_{15}+2\,D_{13})P_{2}(\cos\theta)\,,
\end{eqnarray}
\begin{equation}
3A_{10}=2\,P_{33}+P_{13}-2P_{31}-P_{11}+3(D_{15}-D_{13})\cos\theta\,,
\end{equation}
\begin{equation}
3A_{11}=P_{33}-P_{13}-P_{31}+P_{11}-3(D_{15}-D_{13})\cos\theta\,.
\end{equation}

\begin{center}
{\bf  2. $\pi N \leftrightarrow \eta N$ reaction}
\end{center}

In this case due to the isospin zero nature of the eta meson only total isospin
$I=\frac{1}{2}$ is allowed. Therefore, in accordance with definition (8)
we obtain:
\begin{equation}
f_{l0}^{(\pm)}=0
\qquad{\rm{and}}\qquad
f_{l1}^{(\pm)}=-\frac{1}{\sqrt{3}}f_{l\pm}^{1/2}\,.
\end{equation}
The amplitude $\hat{f}_{\pi,\eta}$ in this process is purely isovector, e.g.
$A_{00}^{\pi,\eta}=A_{10}^{\pi,\eta}=0$. The isovector
amplitudes $A_{01}^{\pi,\eta}$ and $A_{11}^{\pi,\eta}$ are given by
\begin{equation}
\sqrt{3}A_{01}^{\pi,\eta}=-(S_{11}^{\pi,\eta}+P_{11}^{\pi,\eta}\cos\theta+
2\,D_{13}^{\pi,\eta}P_{2}(\cos\theta))\,,
\end{equation}
\begin{equation}
\sqrt{3}A_{11}^{\pi,\eta}=P_{11}^{\pi,\eta}+3\,D_{13}^{\pi,\eta}\cos\theta\,,
\end{equation}
where we have neglected $P_{13}$ and $D_{15}$ contributions.

\begin{center}
{\bf 3. $\eta N$ scattering}
\end{center}

In this case the amplitude $\hat{f}_{\eta,\eta}$ consists of an
isoscalar part only. Therefore,
\begin{equation}
f_{l0}^{(\pm)}=f_{l\pm}^{1/2}
\qquad{\rm{and}}\qquad
f_{l1}^{(\pm)}=0
\end{equation}
and
\begin{equation}
A_{00}^{\eta,\eta}=S_{11}^{\eta,\eta}+P_{11}^{\eta,\eta}\cos\theta+
2\,D_{13}^{\eta,\eta}P_{2}(\cos\theta)\,,
\end{equation}
\begin{equation}
A_{10}^{\eta,\eta}=-(P_{11}^{\eta,\eta}+3\,D_{13}^{\eta,\eta}\cos\theta)\,.
\end{equation}
The expressions for the $S_{11},\,P_{11}$ and $D_{13}$ amplitudes for
$\pi N \leftrightarrow \eta N$ reaction and $\eta N$ scattering are
taken from Ref.\cite{24}.

\newpage


\bibliographystyle{unsrt}

\newpage

\begin{center}
{\bf FIGURE CAPTIONS}
\end{center}

\begin{enumerate}
\item  Differential cross sections for $\pi^+$ ({\bf a}) and $\pi^-$ ({\bf b})
elastic scattering on $^3$He at pion kinetic energies $T_{\pi}$=100, 200 and
295 MeV calculated with three-body (solid curves) and harmonic oscillator
S-shell (dashed curves) wave functions. The dotted curves are the
PWIA results obtained in the S-shell model. Experimental data are from
Ref.\cite{41} ($\bullet$) and Ref.\cite{42}(o).

\item  Differential cross sections for $\pi^+$ elastic scattering on
the deuteron at pion kinetic energies $T_{\pi}$=180--300 MeV calculated with
Paris potential wave function\cite{43}. Solid and dashed curves are full
and PWIA calculations, respectively. Experimental data are from
Ref.\cite{44}.

\item Same as in Fig.1 for pion energies $T_{\pi}$=350, 400,
and 500 MeV. The dash-dotted curves are the results of calculations
in the phenomenological model with C0 form factors extracted from charge
distributions of $^3$He and $^3$H. Experimental data are from Ref.\cite{45}.

\item The target asymmetry $A_y$ in $\pi^+$ ({\bf a}) and $\pi^-$ ({\bf b})
elastic scattering on $^3$He at $T_{\pi}$=100, 295 and 500 MeV. The notations
for the curves are the same as in Fig.1. Experimental data at $T_{\pi}$=100 MeV
are from Ref.\cite{42}.

\item  Nuclear structure effects in the  superratio. Solid curves are the
results obtained using three-body wave functions and  dashed curve - using
S-shell wave functions with harmonic oscillator parameter
$b=1.65 fm$ both for the $^3$He and $^3$H . Dash-dotted curves are
the S-shell model results with $b=1.65 fm$ for $^3$H and
$b=1.68 fm$ for $^3$He. Experimental data are Refs.\cite{48}($\times$),
\cite{49}($\bullet$) and \cite{50}(o).

\item Differential cross sections for $\pi^{\pm}$ elastic scattering on
$^3$He ({\bf a}) and $^3$H ({\bf b}) at $T_{\pi}$=180 MeV. The dashed and solid
curves show our calculations using three-body wave functions with and
without second-order potential from Ref.\cite{8} respectively. Experimental
data are from Ref.\cite{49}($\bullet$) and Ref.\cite{50}(o) (preliminary).

\item Differential cross sections for $^3$H($\pi^+,\pi^0)^3$He at
$T_{\pi}$=130 MeV and $^3$He($\pi^-,\pi^0)^3$H reaction at $T_{\pi}$=200 MeV.
In the latter case the contributions of spin-flip (long-dash-dotted) and
non-spin-flip (long-dashed curve ) parts of $\pi N$ amplitude are shown
separately. The notations of other curves are the same as in Fig.1.
Experimental data for the ($\pi^+,\pi^0$) reaction are from Ref.\cite{53},
data for the ($\pi^-,\pi^0$) reaction are from Ref.\cite{54}($\bullet$) and
Ref.\cite{55}(o).

\item  Differential cross sections for $^3$He($\pi^-,\pi^0)^3$H at
$T_{\pi}$= 285, 428 and 525 MeV calculated in the three-body (solid curves )
and phenomenological (dash-dotted curves ) models. Experimental data are
from Ref.\cite{56}.

\item Phenomenological isovector second-order contribution in SCE and elastic
channels at $T_{\pi}$=290 MeV. Solid and dashed curves are the results without
and with $V^{(2)}$-term respectively. Experimental data are from\cite{53} for
SCE and \cite{41} for elastic scattering.

\item The target asymmetry $A_y$ for $^3$He($\pi^-,\pi^0)^3$H at
$T_{\pi}$= 100, 295 and 500 MeV. The notations are the same as in Fig.1.
The dash-dotted curves are the results of calculations in the phenomenological
model.

\item Differential cross section ({\bf a}) and target asymmetry ({\bf b}) for
$^3$He($\pi^-,\eta)^3$H at $T_{\pi}$=555 MeV. The solid (dashed) curves
are the results of full (PWIA) calculations with three-body wave functions
and the dash-dotted curves are results obtained in the phenomenological
model. The dotted curves are PWIA results obtained with S-shell wave
functions. Experimental data are from Ref.\cite{25}.

\item Eta c.m. momentum dependence for the differential cross section
of $^3$He$(\pi^-,\eta)^3$H at 0$^{\circ}$. The dash-dotted and full curves
show our full calculations using phenomenological and three-body wave
functions, respectively. The dotted curve is obtained with  three-body wave
functions in PWIA. Experimental data are from Ref.\cite{25}.

\end{enumerate}
\end{document}